# Attoliter Mie Void Sensing

Serkan Arslan[1,*], Micha Kappel[1], Adrià Canós Valero[2], Thu Huong T. Tran[1], Julian Karst[1], Philipp Christ[2], Ulrich Hohenester[2], Thomas Weiss[1,2], Harald Giessen[1], and Mario Hentschel[1*]

[1]*4th Physics Institute and Research Center SCoPE, University of Stuttgart, Pfaffenwaldring 57, 70569 Stuttgart, Germany*

[2]*Institute of Physics, University of Graz and Nawi, Universitätsplatz 5, 8010, Graz, Austria*

*serkan.arslan@pi4.uni-stuttgart.de, m.hentschel@physik.uni-stuttgart.de

**Abstract**

Traditional nanophotonic sensing schemes utilize evanescent fields in dielectric or metallic nanoparticles, which confine far-field radiation in dispersive and lossy media. Apart from the lack of a well-defined sensing volume that can be accompanied by moderate sensitivities, these structures suffer from the generally limited access to the modal field, which is key for sensing performance. Recently, a novel strategy for dielectric nanophotonics has been demonstrated, namely, the resonant confinement of light in air. So-called Mie voids created in high-index dielectric host materials support localized resonant modes with exceptional properties. In particular, due to the confinement in air, these structures benefit from the full access to the modal field inside the void. We utilize these Mie voids for refractive index sensing in *single* voids with volumes down to 100 attoliters and sensitivities on the order of 400 nm per refractive index unit. Taking the signal-to-noise ratio of our measurements into account, we demonstrate detection of refractive index changes as small as $6.9 \times 10^{-4}$ in a defined volume of just 850 attoliters. The combination of our Mie void sensor platform with appropriate surface functionalization will even enable specificity to biological or other analytes of interest, as the sensing volumes are on the order of cellular signaling chemicals of single vesicles in cellular synapses.

Sensing applications have been a key driving force behind nanophotonics and plasmonics[1–4] and include medical demands such as the detection of biomolecules and their conformational changes,[5–7] safety-related concerns such as the measurement of gas concentrations, monitoring of chemical reactions,[8] and many more.[9–12] Ultimately, achieving the highest sensitivities for smallest quantities, possibly even down to the single molecule level, requires optimization of the sensing platform.[13–24] Optimization, however, depends on the sensing application.[25–28] While the quality factor of the resonances is important to discriminate spectral shifts, also the mode volume has to be taken into account in order to optimize the sensor performance.[29–34] While exhibiting excellent sensitivity and performance,[35,36] highest quality factors are often observed for extended modes in gratings or periodic structures, resulting in very large sensing volumes. In other cases, high-quality factors are associated with strongly confined modes, hampering interaction with materials of interest.[37] Combining whispering gallery modes resonators with, e.g., plasmonic nanoantennas has been shown to be a powerful alternative approach.[38,39]

Additionally, pushing a sensor to the ultimate limit requires an ultimately small sensing volume. In this letter we introduce Mie voids as a novel sensing platform, making use of the unique full accessibility of their modal fields.[40] Mie voids, which are void structures in high refractive index dielectrics that strongly confine light, are thus ideal candidates for nanoscale optical sensing as they maximize the far-field response by maximizing the analyte-resonator interaction in a well-defined and ultimately small volume.[40,41] The optical response is in fact so large that the change in optical properties of individual resonators can be observed with bare eyes in an optical microscope. Ultimately, we demonstrate that single Mie voids can report refractive index changes as small as 6.9 x $10^{-4}$ inside a volume of only 850 attoliters. Beyond the current experiments, surface functionalization can introduce selectivity to specific analytes and allow for their discrimination. Mie voids are also ideal as a platform for microfluidic sensing chips due to the ease of fabricating voids in silicon substrates and their straightforward integrability into this platform.[42–44]

**Results and Discussion**
Figure 1a illustrates the basic concept of Mie void sensing. The artistic sketch depicts a single Mie void in a gallium arsenide surface. Such a void supports resonant modes by virtue of the finite Fresnel reflection at the interfaces, confining the radiation to the void region which allows full access to the modal field distributions. Mie voids are thus an ideal nanoscale sensor for liquid analytes, serving not only as the analyte container of well-defined volume but also supporting a localized resonant mode and thus ensuring maximized interaction. In the experiment, we collect white light reflection spectra from the structures, and record the changes to the optical response for different analytes. The setup allows us to measure the reflectance spectra of arrays of Mie voids as well as individual ones.

Figure 1b depicts the basic working principle. The leftmost scanning electron microscopy (SEM) images show arrangements of individual Mie voids of varying diameter and nominally identical depth. The top depicts a spiral of voids with continuously increasing diameter, and the bottom one displays a random arrangement of voids with different diameters. The middle and rightmost columns depict optical microscope images of the same arrangement of voids in air and a liquid surrounding of refractive index n = 1.38 (propanol). One can clearly observe the resonant scattering of the individual voids in dependence on their diameter. This manifests as a significant color change, that is, resonance shift for different refractive index surroundings. The images for the air and liquid surroundings reveal a very strong color change from an orange-to-red-dominated hue for air-filled to a purple-to-blue-dominated one for the propanol-filled voids.

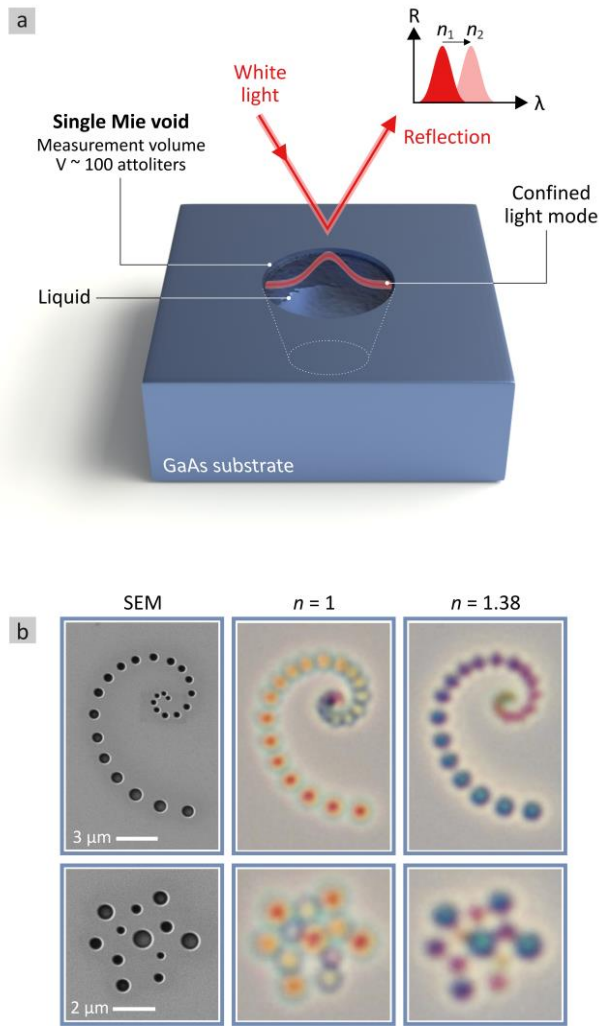

**Figure 1: a** Basic idea of Mie void-based sensing. The well-defined volume of the resonant nanostructure makes a Mie void an ideal nanoscale sensor, maximizing interaction of the analyte with the localized mode inside the void. **b** Scanning electron microscope (SEM) images and optical micrographs of arrangements of individual Mie voids with different sizes and thus different resonance positions. Each void is associated with a distinct color impression. Changing the refractive index of the surrounding from 1 to 1.38 for propanol causes the spectral resonance to shift dramatically, underpinning the working mechanism of the Mie void sensor.

In the course of this discussion, we will progress from array measurements and larger refractive index changes achieved via mixtures of solvents to the ultimate limit at single voids and smallest changes obtained by utilizing the thermo-optical refractive index change of the analyte.

The measurement setup is sketched in Figure 2a. We perform refractive index measurements of liquids inside a flow cell with a source and drain to flush the cell with liquids, in this case, propanol and toluene (n = 1.38 and 1.49, respectively). These values stem from taking the average value of the refractive index dispersion between 500 nm and 1000 nm.[45,46] By choosing an appropriate mixture of the two liquids, we can obtain refractive indices between these two extreme values. In the experiment, we utilize mixing ratios of (10:0, 8:2, 6:4, 4:6, 2:8, 0:10), corresponding to refractive indices of (1.380, 1.402, 1.424, 1.446, 1.468, 1.490). Through a glass window in our cell, the Mie void sample can be imaged with a 60x Nikon objective which is part of a microspectroscopy setup based on an inverse Nikon TE2000-U microscope.[19,47] As the substrate is intransparent, we measure reflectance spectra in a bright-field geometry. For this, we utilize Köhler illumination and collect the reflected light with the help of a 4f setup coupled to a Princeton Instruments grating spectrometer (SP2500i). Both the illumination and collection paths utilize apertures in the Fourier plane (A-stop) in order to limit the illumination and collection to angles close to normal incidence, effectively reducing the NA of the objective. As we measure in bright-field reflectance geometry, we use an aperture in the image plane of the collection path (F-stop) to only collect light from the region of interest (arrays or single voids). For the single void measurements, we utilize fixed-size apertures in the collection path. For the 60x objective, pinholes of 100 µm or 75 µm diameter were used depending on the void size.

Figure 2b presents reflectance spectra for different analyte combinations. The data shown corresponds to three different arrays of voids with different radii as displayed in the SEM images to the left. The void radii are 265 nm, 395 nm, and 570 nm, respectively (top radius) with nominally identical depths at 480 nm. The middle panel shows the reflection spectra in a waterfall plot in dependence of the different analyte refractive indices.

We first consider the general shape of the measured reflectance spectra. Looking at the sweep in the middle corresponding to the medium-sized voids we observe the distinct resonant peak features of the three lowest-order Mie void modes, labeled as modes 1 to 3. In comparison, these modal features are blue-shifted in the spectra of the smaller voids at the bottom, where only modes 1 and 2 are observable. Conversely, in the spectra of the biggest voids at the top, the modes are red-shifted with modes 2 and 3 being clearly visible and parts of the 4[th] mode shifting into the visible range above 400 nm.

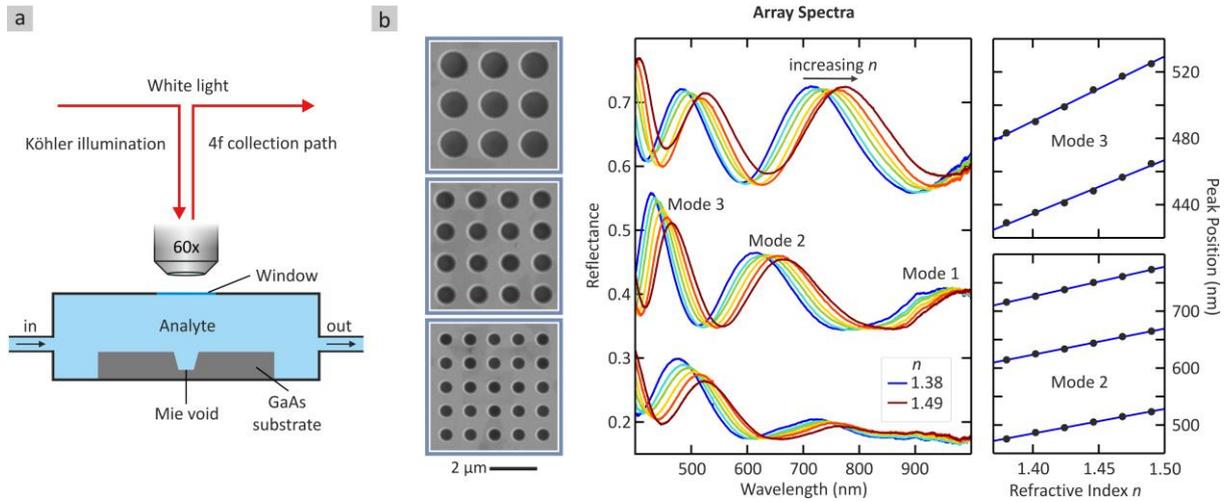

**Figure 2: a** Sketch of the measurement setup consisting of a fluid cell for the exchange of different analytes with different refractive indices as well as the illumination and collection path of the white light reflection measurement. **b** SEM micrographs as well as reflectance spectra for different arrays of Mie voids and the extracted shifts of the modal features. The top radii of the voids are 265 nm, 395 nm, and 570 nm, with a nominally identical depth of 480 nm. The spectra for the two larger radii have been shifted upwards for clarity (by 0.225 and 0.5 respectively).

We now turn our attention to the resonance shifts upon exchange of the analytes. For all three arrays, we observe a clear red shift for increasing refractive indices, as expected. In order to analyze these wavelength shifts in detail, we extract the peak positions from the spectra and plot them over the corresponding refractive index in the right column of Figure 2b. As only the modes 2 and 3 are entirely within the measurable wavelength range, we focus our discussion on these. For all three void sizes, a clear linear dependence between peak position and refractive index can be observed for both modes. This result is intuitive as the refractive index increases the optical path length and therefore the effective resonator size, thus lowering the energy of a given mode. By fitting linear functions, we obtain the sensitivity for the different void sizes and mode orders from these data. We define sensitivity as wavelength shift of peak position per refractive index unit (RIU). For mode 2 from biggest to smallest void the slopes are $\frac{\Delta\lambda}{\Delta n} = (523 \pm 7)\frac{nm}{RIU}$, $\frac{\Delta\lambda}{\Delta n} = (459 \pm 9)\frac{nm}{RIU}$ and $\frac{\Delta\lambda}{\Delta n} = (433 \pm 8)\frac{nm}{RIU}$. For mode 3 the biggest void exhibits a sensitivity of $\frac{\Delta\lambda}{\Delta n} = (391 \pm 10)\frac{nm}{RIU}$ and the medium sized one $\frac{\Delta\lambda}{\Delta n} = (323 \pm 12)\frac{nm}{RIU}$.

One key feature of Mie void sensing is the well-defined and ultra-small sensing volume of the single void, which is only on the order of a few hundred attoliters. This intriguing property can only be fully exploited when addressing a single Mie void. Results in Figure 1b already indicate that individual voids are clearly observable in a microscope with naked eye, thus it appears more than feasible to also extract the spectral information.

For these measurements, we have chosen the Mie void depicted in the SEM image in Figure 3a with a volume of only ~370 attoliters (550 nm top radius, 435 nm bottom radius, 480 nm depth). For this individual Mie void, Figure 3b displays the reflectance spectra taken for the same set of refractive indices as before. Despite the fact that we are addressing a single Mie void and thus interrogate only an analyte volume of 370 attoliters, we observe well-modulated spectra with an excellent signal-to-noise ratio. As expected, we observe a red shift of the spectra for increasing refractive indices. We extract the peak position for each refractive index step and plot it as a function of the corresponding refractive index in Figure 3c, which shows the expected linear dependence. The fit returns a sensitivity of $\frac{\Delta\lambda}{\Delta n} = (399 \pm 7)\frac{nm}{RIU}$. We thus observe similar sensitivities compared to the array measurements, corroborating the fact that voids only interact very weakly with each other.

In order to further underpin our results, we performed full-wave simulations with the Finite Element Method implemented in COMSOL Multiphysics, (Figures 3d and 3e). Figure 3d displays the absolute value of the electric field at the resonance wavelength. Three field maxima can be observed within the void, confirming the excitation of a high order mode of the resonator, in contrast with the low order modes investigated previously.[40] The simulated reflectance spectra are depicted in Figure 3e and show good agreement with the measurements. Figure 3f depicts the simulated shift of the peak position against the corresponding refractive index. The sensitivity of $\frac{\Delta\lambda}{\Delta n} = (381 \pm 4)\frac{nm}{RIU}$ is in excellent agreement with the measurement.

Another important cross-check is related to the actual active sensing volume. In our experiment not only the void itself but also the volume above the sample is filled with the analyte. In order to investigate its impact, we performed simulations assuming only the void filled, void and space above filled, and an empty void with filled space above (see Supporting Information Figure S1). The simulations clearly prove that a spectral shift of the features is only observed for an analyte-filled void and is barely influenced by the analyte above the sample.

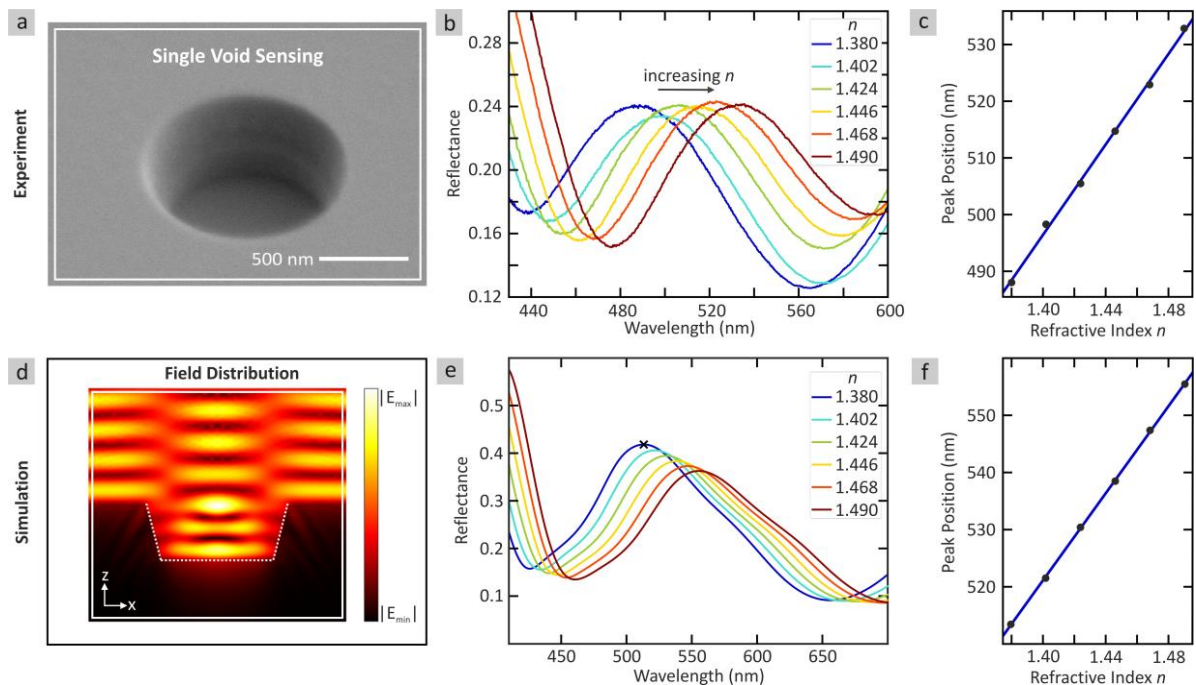

**Figure 3:** Single Mie void refractive index sensing. **a** 45° tilted-view SEM image of the used Mie void. **b** Reflectance spectra for different analyte solutions. **c** Extracted peak position with respect to the corresponding refractive index **d** Simulated electric field intensity distribution for a void filled with analyte of refractive index 1.38 at a wavelength of 513 nm (resonance peak). **e** Simulated single Mie void reflectance spectra. **f** Extracted peak positions for the simulated spectra.

The sensing volume can be further decreased by using smaller-sized Mie voids. To illustrate this, Figure S2 in the Supporting Information presents measurements for a Mie void with a volume of only 100 attoliters, still offering good sensing performance with a sensitivity of $\frac{\Delta\lambda}{\Delta n} = (516 \pm 13)\frac{nm}{RIU}$. Two constraints must be considered here: First, the resonances will blue-shift, at ultimate small sizes toward the UV spectral range, which is outside of the easy-to-access spectral range. Second, the coupling of the modes with the far field will decrease due to the decreased volume and thus diminish the observable signal-to-noise ratio.

Utilizing a sensing platform with ultimate small sensing volumes on the order of a few hundred attoliters becomes particularly interesting when pushing it simultaneously to the smallest detectable refractive index changes, exploring the overall limits of the platform. In order to investigate this behaviour, very small refractive index changes have to be realized in a controlled and reproducible way. In our experiments, we make use of the thermo-optic effect, that is, the dependence of the refractive index of a material on its temperature. In a setup as sketched in Figure 4b, setting a certain temperature, controlled by a PID feedback loop, allows us to very accurately choose a certain refractive index and thus also allows to finely tune this refractive index in ultimately small steps. To this end, we utilized a cell sketched in Figure 4a. A droplet of glycerol was trapped between a cover slip and a Mie void sample with a few layers of capton tape acting as a spacer and rim at the edges, sealed with glue. Glycerol as an analyte was chosen for its large thermo-optic coefficient.

We performed single Mie void spectral measurements analogous to the previous measurements but now as a function of temperature, from which we can determine the corresponding refractive index via the thermoptic coefficient as described in the supporting information.[48] The utilized Mie void displayed in the SEM image of Figure 4 c has a top radius of 775 nm, a bottom radius of 465 nm, a depth of 685 nm, and a volume of 850 attoliters. For this single void, we measure spectra from 30°C to 55°C in steps of 5K. For the extreme cases of 30°C and 55°C, the reflectance spectra around 635 nm are shown in Figure 4d. Even for the largest refractive index shift, which corresponds to 5.75 x $10^{-3}$, the spectral shift is barely visible with the naked eye. In order to extract the spectral shift from the measured spectra we thus utilized the so-called centroid method.[20] This method is robust against noise as it takes a larger spectral region into account and thus generally more suitable for detecting very small spectral shifts.

In our measurement presented in Figure 4e we cycled between 30°C and 55°C five times (three times up, two times down) and extracted the centroid positions, resulting in three data points at the maximum and minimum temperature and five data points each for temperature setting in-between. The refractive index values on the upper horizontal axis were calculated via the thermo-optic coefficient of glycerol and the applied temperature. Note that the individual steps correspond to a refractive index change of just 1.15 x $10^{-3}$.

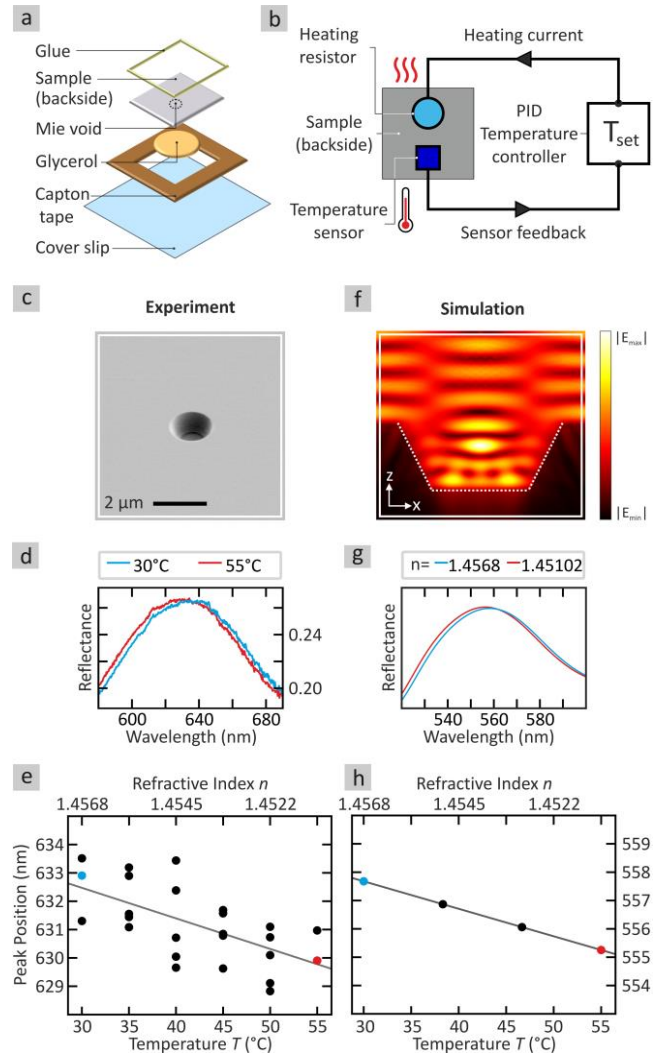

**Figure 4:** Ultimate limits of single Mie void-based refractive index sensing. **a** Sketch of the measurement cell. **b** Measurement scheme utilizing the thermo-optic coefficient of glycerol. Via a PID temperature controller the sample and analyte temperature and thus the refractive index can be precisely controlled. **c** SEM micrograph of the measured Mie void. **d** Reflectance spectrum at the extreme temperature points of 30° and 55°. **e** Extracted peak position versus the applied temperature and corresponding refractive index for a five-cycle measurement run. **f** Simulated electric field intensity distribution. **g** Simulated reflectance spectra corresponding to the experimental data in d. **h** Simulated peak shift for the refractive index range covered in the experimental data in e.

The measurement clearly shows that the change in refractive index is imprinted on the optical properties and can thus be extracted. A blue shift in resonance position for increasing temperature is still observable, which matches the negative sign of the thermo-optic coefficient of glycerol.[48] The variation in the obtained spectral positions demonstrates that we are very close to the detection limit of our sensing platform. Yet, we are still able to resolve the extremely small changes. It should be noted that due to the multiple measurements we can indeed extract a refractive index change on the order of $6.9 \times 10^{-4}$ inside a volume of just 850 attoliters. The slope of a linear fit to the data points yields a sensitivity of $\frac{\Delta\lambda}{\Delta n} = (470 \pm 125)\frac{nm}{RIU}$, comparable to the earlier measurements. While the sensing limit is smaller compared to the case of microinterferometry (about $10^{-8}$ RIU at $10^{-9}$ liters), also exhibiting a well-defined analyte volume, Mie voids show a significantly smaller overall sensing volume (6-7 orders of magnitude).[49] Additionally we explain in the supporting information S3 that the sensitivity of the measurement can even be increased further by using Mie void arrays instead of individual voids, effectively giving a trade-off between sensitivity and minimal sensing volume.

Figure 4f depicts the calculated electric field intensity distribution at resonance. The results reveal an even more complicated field distribution in comparison with Figure 3. Indeed, as expected from the size of the void, we utilized an even higher order mode in these experiments, aiding the sensitivity. In Figure 4g, we plottted the simulated reflectance spectra for refractive indices of 1.4568 and 1.45102 (corresponding to 30 and 55°C, respectively). In Figure 4h, we have drawn the simulated resonance shift over the refractive index, obtaining a sensitivity of $\frac{\Delta\lambda}{\Delta n} = 422 \frac{nm}{RIU}$. Overall, the simulation is in excellent agreement with the experiment.

In summary we have demonstrated the use of Mie voids as sensing platform, making use of the full access to the modal field distribution as well as the well-defined and ultra-small sensing volume on the order to a few hundred attoliters. Extended mode sensing technique show sizeably higher sensitivities, yet, have significantly larger or not such a well-defined sensing volume. Taking the sensing volume as well as the measured sensitivity into account, single Mie void sensing outperforms traditional techniques with well-defined sensing volumes, such as microinterferometry.[49] Our ansatz is therefore ideal for the interrogation of ultra-small analyte and material volumes, which can be straightforwardly filled into the cuvette, conveniently formed by the Mie void itself. In the future, functionalization can also add specificity,[42,50–53] moving beyond proof-of-concept refractive index sensing. Particularly interesting in our case is the incorporation of hydrogels and similar material classes into the void, again making use of the intrinsic full access to the field and therefore maximising interaction and in turn sensitivity.[54] Also, the combination with microfluidics and the read out using a cheap CMOS color camera could render our ansatz very attractive for automated pharmaceutical and chemical analytics.

**Acknowledgments**

This work was supported the Ministerium für Wissenschaft, Forschung und Kunst Baden-Württemberg (RiSC Project "Mie Voids", ZAQuant), Vector Stiftung MINT-Innovationen, Baden-Württemberg-Stiftung (Opterial), European Research Council (ERC Advanced Grant Complexplas & ERC PoC Grant 3DPrintedOptics), Bundesministerium für Bildung und Forschung, Deutsche Forschungsgemeinschaft, (SPP1839 "Tailored Disorder" and GRK2642 "Towards Graduate Experts in Photonic Quantum Technologies").